\documentclass[review]{elsarticle}

\usepackage{lineno,hyperref}     
\modulolinenumbers[5]            

\journal{Journal of Molecular Spectroscopy}

\usepackage{braket}	
\usepackage{mhchem}	

\usepackage{kbordermatrix}	
\usepackage{amsmath}	

\usepackage{graphicx}	
\usepackage{multirow}	
\usepackage{wrapfig}	
\usepackage{subcaption} 
\usepackage{dcolumn}	

\usepackage{comment}	
\usepackage{verbatim}	
\usepackage[normalem]{ulem}	
\usepackage{lineno}	

\usepackage{xcolor}	
\usepackage{url}	

\graphicspath{ {./figures/} }		
\newcommand{\wn}{cm$^{-1}\!\!$ }		
\newcommand{\us}{$\mu$sec }			

\newcommand{\vib}[2]{\textit{v}$_#1 = #2$}	

\newcommand{\APi}{$\tilde{A}\,^2\Pi$}
\newcommand{\XSig}{$\tilde{X}\,^2\Sigma^+$}

\begin{document}

\begin{frontmatter}

\title{The 1.66\,$\mu$m Spectrum of the Ethynyl Radical, CCH}

\author{Eisen C. Gross}
\ead{eisen.gross@stonybrook.edu}
\address{\footnotesize{Chemistry Department, Stony Brook University, Stony Brook, NY 11794-3400}}
\author{Anh. T. Le}
\ead{anh.t.le@asu.edu}
\address{\footnotesize{School of Molecular Sciences, Arizona State University, Tempe, AZ 85287}}
\author{Gregory E. Hall}
\ead{gehall@bnl.gov}
\address{\footnotesize{Chemistry Division, Brookhaven National Laboratory, Upton, NY 11973-5000}}
\author{Trevor J. Sears\corref{mycorrespondingauthor}}
\ead{trevor.sears@stonybrook.edu}
\address{\footnotesize{Chemistry Department, Stony Brook University, Stony Brook, NY 11794-3400}}
\cortext[mycorrespondingauthor]{Corresponding author}

\begin{abstract}
Frequency-modulated diode laser transient absorption spectra of the ethynyl radical have been recorded at wavelengths close to 1.66 $\mu$m.  The observed spectrum includes strong, regular, line patterns.  The two main bands observed originate in the ground \XSig state and its first excited bending vibrational level of $^2\Pi$ symmetry. The upper states, of $^2\Sigma^+$ symmetry at 6055.6 \wn and $^2\Pi$ symmetry at 6413.5 \wn, respectively, had not previously been observed and the data were analyzed in terms of an effective Hamiltonian representing their rotational and fine structure levels to derive parameters which can be used to calculate rotational levels up to J = 37/2 for the $^2\Pi$ level and J = 29/2 for the $^2\Sigma$ one.  Additionally, a weaker series of lines have been assigned to absorption from the second excited bending, (020), level of $^2\Sigma$ symmetry, to a previously observed state of $^2\Pi$ symmetry near 6819 \wn. These strong absorption bands at convenient near-IR laser wavelengths will be useful for monitoring CCH radicals in chemical systems.
\end{abstract}


\end{frontmatter}

\date{\today}

\section{Introduction} \label{Intro}
The ethynyl radical plays an important role as a chemical intermediate in practically any chemical environment containing energetic hydrocarbon species.  It also provides a textbook example of the quantum mechanical energy level complexities resulting from multistate vibronic coupling at low and experimentally accessible internal excitation energies. As a consequence, CCH has been the subject of numerous experimental and theoretical studies during the past 20 years. The excited \APi\, state, resulting from a promotion of a bonding $\pi-$ electron to the non-bonding $\sigma-$ orbital in a single-configuration picture, lies only some 3600 \wn above the \XSig ground state in the radical.\cite{Sharp-Williams2011a} This energy is practically the same as that for excitation of of the C-H stretching vibration and the levels are strongly mixed, as are many other vibrationally excited ground state levels that extend into the $\tilde{A}$-state energy region.  \\

The infrared and near-infrared spectrum of the radical \cite{Forney1995} was observed to be strong and extensive, but detailed spectroscopic assignments were initially elusive. The results of calculations by Tarroni and Carter\cite{Tarroni2003,Tarroni2004} provided the quantitative understanding that allowed the earlier spectroscopic vibronic level observations to be built into a rational overall picture and references in these papers summarize the earlier work. Tarroni and Carter showed that even the lowest excited bending level of ground state CCH has significant \APi \,wavefunction character and all higher vibronic levels possess mixed electronic character.  Their results also showed that the strong and extensive infrared and near-infrared spectrum derived from the \APi$-$\XSig band system are built on a strong progression in the C-C stretching vibration whose intensities are shared among multiple interacting levels.  \\

While the computational results led to an understanding of the vibronic energy levels of the radical, the rotational, fine and even hyperfine structure associated with them has subsequently been an active area of study. The interest lies not only in mapping the complexity and providing spectroscopic avenues for monitoring the radical concentration in, for example, the chemistry of C$_2$H with hydrocarbon molecules,\cite{Kovacs2010} in soot formation\cite{Frenklach2002, Appel2000, Wang2011, Frenklach2020} and in the production of carbon nanotubes,\cite{Wang2014} but also in characterizing state couplings\cite{Sharp-Williams2011a,Sharp-Williams2011b, Tokaryk2015} and the kinetics of vibronic relaxation.\cite{Le2016}   \\

Herein, we report new high resolution spectra of the radical in a previously unexplored region of the near-infrared in the gas phase.  The strong bands observed involve the ground state and its first excited bending level and they will be useful for future analytical, kinetics and dynamics studies of this important radical at wavelengths convenient for inexpensive diode and fiber lasers.

\section{Experimental Details} \label{Expt}
\begin{figure}[h]
    \begin{center}
       \includegraphics[width=13cm]{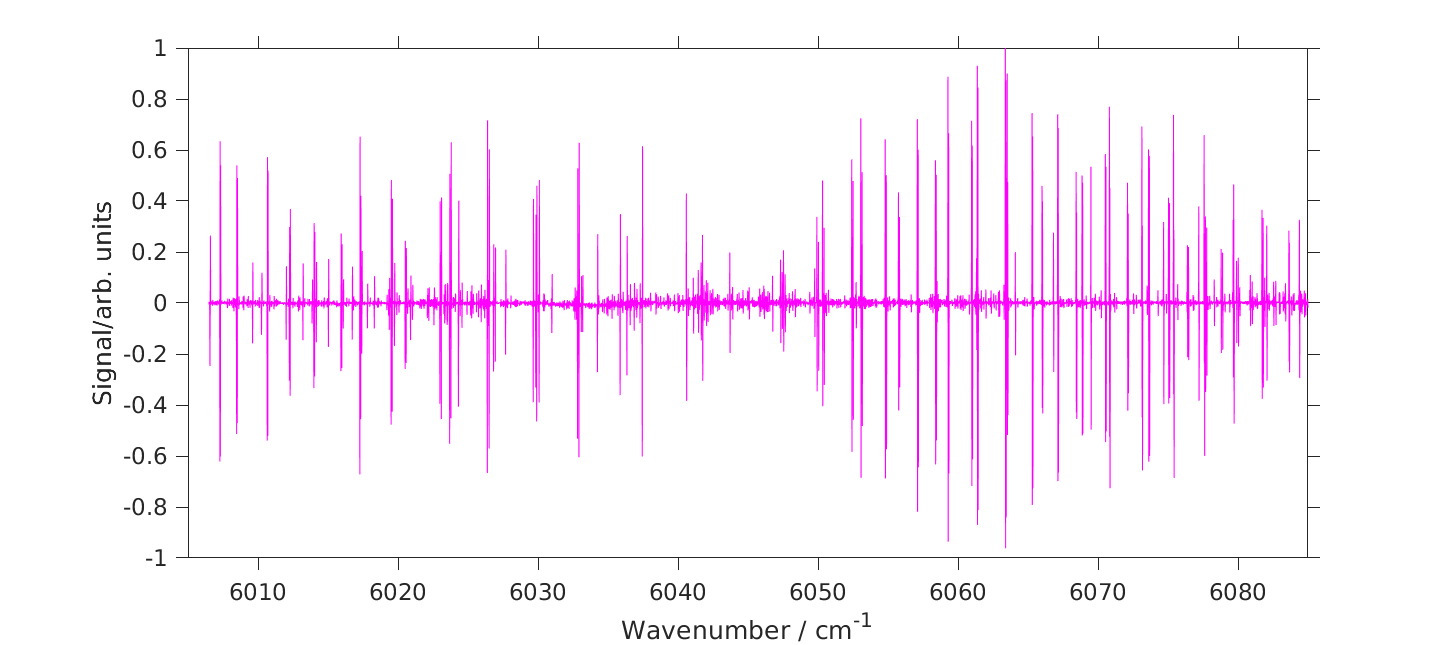}
    \end{center}
    \caption{\label{fig:6005_6085}\footnotesize{Overview of the observed spectrum in the 6040 \wn region. Absorption lines appear as the first derivative due to the FM detection scheme used.}}
\end{figure}

  The data were recorded several years ago at Brookhaven National Laboratory, and the frequency-modulated laser transient absorption spectrometer used has been described in detail previously.\cite{Le2018, Le2016, Chang2011, Forthomme2012, Forthomme2013}  The radicals were formed by a 193 nm \ce{ArF} excimer laser photolysis of a sample of (1,1,1)-trifluoro propyne, \ce{CF3CCH}.  Premixed samples containing 10 \% of precursor diluted in argon were prepared and the mixture flowed slowly through the absorption cell at a constant pressure of 500 mtorr.  The absorption cell included Herriott type multipass optics for the probe laser beam.  The active pathlength was several meters.\cite{Le2016, Le2018}  The \ce{ArF} photolysis laser firing pulse was used to trigger the digital oscilloscope used to acquire the data.  A Bristol Instruments model 621B wavemeter was used to measure the diode laser wavenumber.  During our scans, the diode laser would remain at a given frequency for 40 photolysis shots, the signals from which were averaged and saved, before the diode laser stepped to the next frequency position.  The excimer laser's repetition rate was 10 Hz, and pulse energies typically 100 mJ/pulse, although only $\sim40$ mJ/pulse reached the cell.  To maintain the energy inside the cell, it was necessary to clean the cell's windows between every 3 spectroscopic scans of approximately 0.8 \wn \, in length.  An overview of the spectrum obtained is shown in Figure \ref{fig:6005_6085}. \\

\section{Results and Analysis}
\subsection{Introduction}

Multiple series of strong transient absorption signals were observed in the region from about 6005 to 6085 \wn.   One series consisted of a clear, repeating pattern of 4 lines, suggesting the splitting pattern of a spin doublet radical, combined with a non-zero orbital angular momentum.  Another set of two series of single lines fell into clear P- and R-branch patterns. \\

\subsection{The $(6413.5)^2\Pi\,\leftarrow \,(010)^2\Pi$ transition}
For the series consisting of patterns of 4 lines, initial assignments were made from lower state combination differences calculated using the rotational and fine-structure levels of the $(010)\,^2\Pi$ excited bending vibrational level of the radical. The molecular constants for the this level are well known from work by many previous authors summarized in our previous paper.\cite{Le2018}  They are given in table \ref{tab:lower_parms}. Note that table I of reference \cite{Le2018} has a typographical error in the sign of the spin-orbit constant $A$, which should be negative. Lower state combination differences were easily identified in the observed spectrum and the upper state was found to also be of $^2\Pi$ character.  The rotational and fine structure patterns in the spectrum are therefore due to a parallel $^2\Pi\, -\, ^2\Pi$ band which has six main branches: $P-$, $Q-$, and $R-$branches for each fine structure component, where each of these is split into two due to the $\Lambda-$ or parity-doubling.  Weaker, cross fine structure transitions, may also be observed for low rotational levels. \\

\begin{table}[t]
		\centering
		\caption{Spectroscopic parameters in wavenumbers (\wn) for the lower states in the observed C$_2$H bands$^a$}
	\label{tab:lower_parms}
{\begin{tabular}{llll}
		\hline
		\hline
		                & $\tilde{X}(0,0,0)$  & $\tilde{X}(0,1^1,0)$ & $\tilde{X}(0,2^0,0)$ \\
		Parameter\,\,   & $^{2}\Sigma ^{+\,\,b}$ &  $^{2}\Pi\,^{c}$ & $^{2}\Sigma ^{+\,\,c}$	\\
		\hline
		$A$                          &                 & -0.3466229(78)   &  \\
		$B$                          & 1.456825742(1) & 1.451285679(25) & 1.45261188(27) \\
		$10^5\,D$                    & 0.35642(12)      & 0.3822(12)    & 0.2457(48)    \\
		$10^6\,H$                    &                  &               &  0.2328(25)$^e$  \\
		$10^2\, \gamma $             & -0.2087934(40)   & -0.115589(20) & -0.116129(59)  \\
		$10^6\, \gamma_{D}$          & -0.165(12)       &             & -0.526(30)     \\
		$10^3\, p$                   &                 & -0.1452(13)    &   \\
		$10\, q$                     &                 & -0.1135841(31) &    \\
		$10^7\,q_D$                  &                 &  0.74(18)      &   \\
        Origin                       &                 &  371.6034(3)$^d$  & 794.300(6)$^f$  \\
		\hline
		\hline
		\multicolumn{4}{l}{\textit{a.} The numbers in parenthesis are one standard deviation}\\
		\multicolumn{4}{l}{ of the fit in units of the last quoted decimal place.}\\
        \multicolumn{4}{l}{\textit{b.} From Reference \cite{Le2018}}  \\
        \multicolumn{4}{l}{\textit{c.} This work, see text for details. }  \\
        \multicolumn{4}{l}{\textit{d.} From Reference \cite{Kanamori1988}} \\
        \multicolumn{4}{l}{\textit{e.} {Also, $L\,=\,-5.29(1)\times10^{-10}$}. Hyperfine parameters } \\
        \multicolumn{4}{l}{$b_F\,=\,30.889(15)$ and $c\,=\,12.821(31)$ in MHz.} \\
        \multicolumn{4}{l}{\textit{f.} From Reference \cite{Chiang1999}} \\
        \\	
	\end{tabular}}
\end{table}
The sum of these splittings lead to the patterns of 4 lines in the prominent $P-$ and $R-$branches.  The $Q-$branches are weaker, decay in intensity quickly with increasing rotational quantum number, but are visible in the spectrum between 6039 and 6042 \wn. \\ 

The complete dataset, consisting of 127 assigned lines, is available in supplementary data for this paper.\cite{Supplemental}  Since the new data contain high precision measurements of fine structure-resolved lower state combination differences, the fit of the $(010)^2\Pi$ lower state rotational intervals from Le et al.\cite{Le2018} was repeated with 12 additional combination differences from the present data.  The final rotational, $\Lambda-$doubling  and fine structure parameters so derived are given in table \ref{tab:lower_parms}. These will be good for computing energy levels up to at least $J = 23/2$, the highest rotational level included in the combination differences data, but likely higher as the present spectral assignments extend to $J = 37/2$.  The hyperfine parameters derived from microwave data\cite{Woodward1987} in the combined fit, were unchanged from those determined by Le et al.\cite{Le2018}   With the lower state energies known and fixed, the new data were fit to the upper state energy levels  using a Matlab script that set up the $^2\Pi$ Hamiltonian operator in the $\mathbf{N}^2$ representation in a parity conserving, case (a)  basis set,\cite{Tokaryk2015,BandC2003}  as given in equation (\ref{equ:paritybasis}).

\begin{equation}\label{equ:paritybasis}
 \ket{\eta J \Omega; \Lambda;S \Sigma (\pm)}\,=\, \frac{1}{\sqrt{2}}\left[\ket{\eta J, \Omega;\Lambda; S, \Sigma}\,\pm\, (-1)^{J-S}\ket{\eta J, -\Omega;-\Lambda; S, -\Sigma}\right]
\end{equation}
Here, $\eta$ represents all the other quantum numbers needed to define the state, $J$ is the total angular momentum excluding nuclear spins, $\Omega$ its projection on the internuclear axis, $\Lambda$, the projection of the orbital angular momentum, and $S$ and $\Sigma$ the electron spin and its projection. In the parity-conserving basis, the projection quantum number labels take only positive values. \\

The $^2\Pi$ Hamiltonian matrix including the higher order centrifugal distortion corrections needed to fit the present data is given in equations (\ref{equ:DPiMtx}) and (\ref{equ:DPiPty}).
\begin{equation}
 \kbordermatrix{ &\ket{J,\frac{1}{2};(\pm)}& \ket{J,\frac{3}{2};(\pm)}\\
  \bra{J,\frac{1}{2};(\pm)} & -A/2 + B(X+1) -\gamma -D X(X+3)&   h.c. \\
                       & + HX(X^2+6X+1)  -\frac{1}{2}(X+3)\gamma_D                    &  \\
  \bra{J,\frac{3}{2};(\pm)} &  -(X-1)^{1/2}(B -\gamma/2 - 2DX   & A/2 + B(X-1) -  \\
                       &   +HX(3X+1)) -(X-1)^{1/2}X\gamma_D                  & D X(X-1) + HX(X^2-1) \\
                       &                                          &+\frac{1}{2}(X-1)^{3/2}\gamma_D
 }\label{equ:DPiMtx}
\end{equation}
Here, the basis functions, equation (\ref{equ:paritybasis}), are represented by $\ket{J,\Omega;(\pm)}$, $X=(J+1/2)^2$ and $h.c$ represents the Hermitian conjugate. Where needed, the average of the off-diagonal elements have been taken to make the matrix symmetric. The parity-dependent terms are:

\begin{equation}
 \kbordermatrix{ &\ket{J,\frac{1}{2};(\pm)} & \ket{J,\frac{3}{2};(\pm)} \\
  \bra{J,\frac{1}{2};(\pm)} & \mp (-1)^{J-S} \frac{1}{2} Y((p+2q)+ (p+2q)_D(X+1)  &      h.c.   \\
                       & +(X-1)Y q_D )                        &    \\
  \bra{J,\frac{3}{2};(\pm)} &  \pm (-1)^{J-S} \frac{1}{2}(X-1)^{1/2} Y(q+\frac{1}{2}(p+2q)_D  &  \mp(-1)^{J-S}(X-1)Y q_D/2  \\
                       & + (X-1)^{1/2} Y^3 q_D)  &
 } \label{equ:DPiPty}
\end{equation}

With $Y\,=\,X^{1/2}$.  The parity-conserving labels $(\pm)$ are related to $e$ and $f$ often used to label the $\Lambda-$doublet components by the convention that levels with $\pm(-1)^{J-S} = +1$ are labeled $e$ and those with $\pm(-1)^{J-S} = -1$ are labeled $f$.\cite{BandC2003}  Electric dipole allowed transitions always connect states of different parity, while the $e/f$ changes depend on the change in the angular momentum $J$ in the transition.\cite{BandC2003}   The molecular constants appearing in equations (\ref{equ:DPiMtx}) and (\ref{equ:DPiPty}) have their usual meanings.  $A$ and $B$ are the spin-orbit and rotational constant, $D$ and $H$ the quartic and sextic centrifugal distortion constants, $\gamma$ and $\gamma_D$ the spin-rotation constant and its centrifugal distortion correction, and  $p$, $q$, $p_D$ and $q_D$ are the $\Lambda$-doubling parameters and their centrifugal distortion corrections.  \\
\begin{table}[t]
		\centering
		\caption{Spectroscopic parameters in wavenumbers (\wn) for the upper states in the observed C$_2$H bands$^a$}
	\label{tab:upper_parms}
{\begin{tabular}{llll}
		\hline
		\hline
		 Parameter  & $^{2}\Sigma ^{+}$ (6055.6)&  $^{2}\Pi$ (6413.5)& $^{2}\Pi$ (6817.9)	\\
		\hline
		$A$                          &                 & -0.8267(68) & -9.3520(30) \\
		$B$                          & 1.433485(38))   & 1.413930(74) &  1.372000(40)\\
		$10^5\,D$                    & 0.1491(19)      & -0.336(60)    &  -1.498(37)  \\
		$10^7\,H$                    &                 & -0.130(13)     & -0.0445(69) \\
		$10^2\, \gamma $             & -0.211(16)      & 0.121(21)      & 1.974(54) \\
		$10^4\,\gamma_D$             &                 &              & -0.896(25) \\
		$10^3\, p$                   &                 &              &  0.63(36) \\
		$10\,\, q$                   &                 & -0.40100(43) &  0.07254(16) \\
        $10^4\,p_D$                  &                 &              &  0.465(24) \\
		$10^4\,q_D$                  &                 & 0.2067(18)   &       \\
        Origin                       & 6055.6467(14)  & 6413.5257(21) & 6817.8976(11)  \\
        Fit RMS Dev.$^b$             & 2.60           & 4.83          & 1.91   \\
		\hline
		\hline
		\multicolumn{4}{l}{\textit{a.} This work. The numbers in parenthesis are one }\\
		\multicolumn{4}{l}{standard  deviation of the fits in units of the last}\\
		\multicolumn{4}{l}{ quoted decimal place.} \\
		\multicolumn{4}{l}{ \textit{b.} Standard deviation of the fit relative to}\\
		\multicolumn{4}{l}{ estimated measurement uncertainty of 0.002 \wn}
        \\	
	\end{tabular}}
\end{table}

The assigned lines and the detailed results of the fitting are available in spreadsheets of supplementary data\cite{Supplemental} for this paper. The overall root mean square standard deviation of the fit to the upper state constants was 4.83, \textit{i.e.} some what lager than the expected measurement uncertainty of approximately $\pm$ 0.002 \wn.  Adding additional flexibility to the fit by including higher order distortion parameters did not appreciably improve the situation, so we conclude the upper state levels suffer from some perturbations. However, no systematic trends are evident in the fit residuals.  The origin of the upper state in the transition was found to be at 6413.5 \wn \, and the rotational and fine structure constants describing the energy levels of the upper state are given in the third column of table \ref{tab:upper_parms}.   

\subsection{The $(6055.6) \, ^2\Sigma \,\leftarrow \, (000)\,^2\Sigma$ transition}
After accounting for most of the strong lines in the observed spectrum, there remained two series of single lines that formed recognizable $P-$ and $R-$branches. A simple combination difference analysis showed that the lower state in this transition was the ground state of the radical, and the spectrum consists of $P-$ and $R-$ branches of the $^2\Sigma\, -\, ^2\Sigma$ transition with unresolved fine structure splittings.  Analysis followed the same procedures as above.  The lower state of the transition is well-known. The rotational and fine structure parameters are given in the second column of table \ref{tab:lower_parms} from reference \cite{Le2018} converted to \wn.  The upper state had not been experimentally observed previously, but Forney et al. \cite{Forney1995} had reported weak lines in this region from samples trapped in low temperature matrices and calculations by Tarroni and Carter \cite{Tarroni2003, Tarroni2004} found a $^2\Sigma$ state at 6054.1 \wn that is likely to be the upper state of the transition observed here.\\

Again, the energy calculation was set up in a case(a) parity-conserving basis set.  The Hamiltonian matrix elements for a $^2\Sigma$ state in this representation are:

\begin{align}
  \bra{\eta J \Omega S \Sigma (\pm)}H\ket{\eta J \Omega S \Sigma (\pm)}\,&=\,BX-D(X^2+X)-(\gamma/2 + \gamma_DX) \notag\\
  &+ H(X^3 + 3 X^2) + L (X^4 + 6 X^3 + X^2) \\
\intertext{The parity-dependent part is:}
  \bra{\eta J \Omega S \Sigma (\pm)}H\ket{\eta J \Omega S \Sigma (\pm)}\,&=\,(\pm)(-1)^{J-S}(-B Y + 2DX+\frac{Y}{2}(\gamma + \gamma_D(1+X)) \notag \\
  &-\,H X Y(3X+1) - 4 L X Y (X^2 + X)),
\end{align}
for levels of $\pm$ parity. Here, $L$ is the octic centrifugal distortion constant, other parameter symbols have the same meaning as above. The energy of a given fine-structure split $^2\Sigma$ rotationally level is the appropriate sum of these two terms, and thus it is completely specified by the rotational quantum number $J$ and the $(\pm)$ parity label. While setting up the calculation in this basis may seem counter intuitive because $^2\Sigma$ states of light molecules such as CCH more closely conform to case(b) coupling,\cite{BandC2003} it is most useful for computing the relative intensities of transitions in perpendicular bands that were important in searching for additional transitions discussed below.  \\

\begin{figure}[h]
  \begin{center}
    \includegraphics[width=13cm]{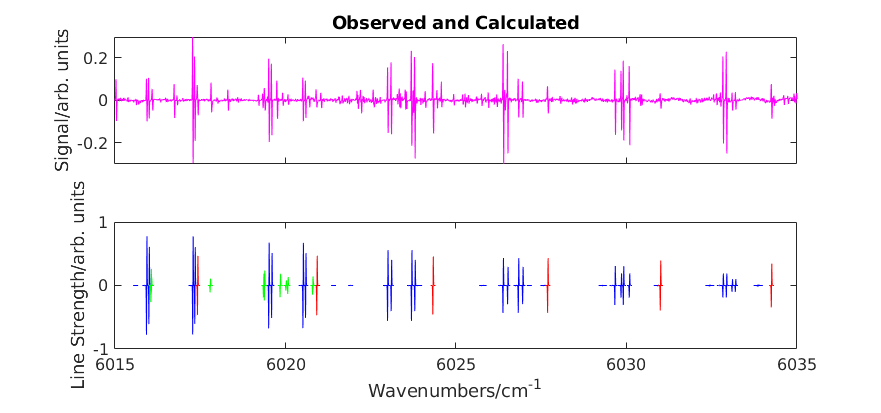}
  \end{center}
    \caption{\label{fig:6015_45}\footnotesize{Section of the observed (upper panel) and calculated spectra in the 6015-6035 \wn \,region, showing contributions from the $^2\Pi$(6413.5)\,-$^2\Pi$(010) band in blue, the $^2\Sigma$(6055.6)\,-$^2\Sigma^+$(000) band in red and weak assigned lines of the $^2\Pi$(6817.9)\,-$^2\Sigma$(020) band in green in the lower panel.}}
\end{figure}

The upper state constants determined in the fitting are given in the second column of table \ref{tab:upper_parms}.  The upper state origin was found to be 6055.6 \wn and the rotational constant 1.4335 \wn.\, Both these quantities are in accord with those predicted by Tarroni and Carter \cite{Tarroni2003} for the level calculated at 6054.1 \wn, which they identified as having the predominant character of a bending overtone, thus confirming the assignment suggested above.  Figure \ref{fig:6015_45} shows an expanded section of the observed spectrum (upper panel) compared to the calculated spectrum that has contributions from both the bands discussed above.  The positions of the calculated lines match the observations well.  There are some inconsistencies between the observed and calculated relative intensities, but these are likely due to the data being recorded at different times with varying degrees of opaque photolysis product build up on the cell window used for the photolysis laser.

\subsection{Kinetics of new bands}
\begin{figure*}[h]
  \begin{subfigure}[h]{0.5\textwidth}
   \centering
   \includegraphics[height=7.0cm]{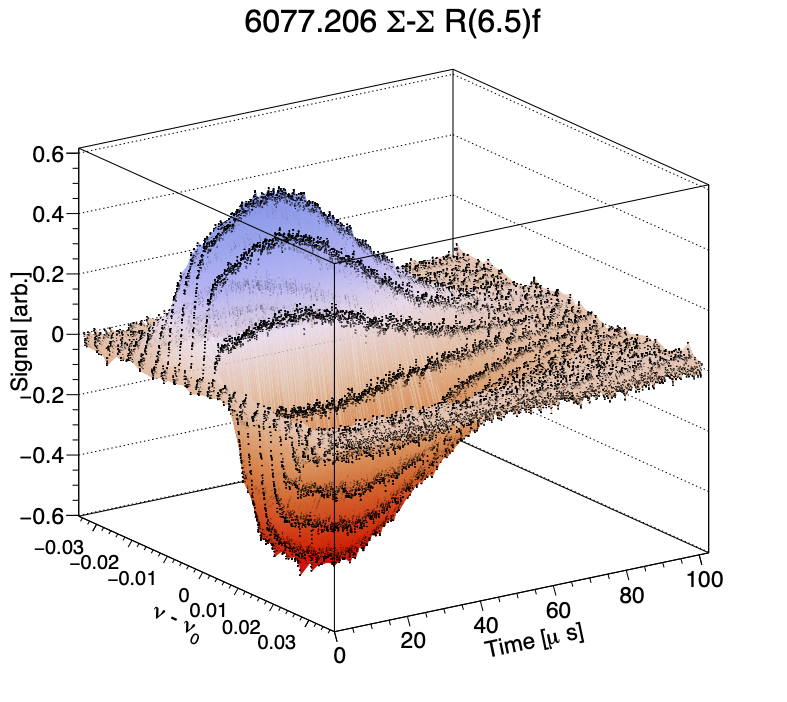}
    \caption{\footnotesize{}}
  \end{subfigure}%
  ~
  \begin{subfigure}[h]{0.5\textwidth}
   \centering
   \includegraphics[height=7.0cm]{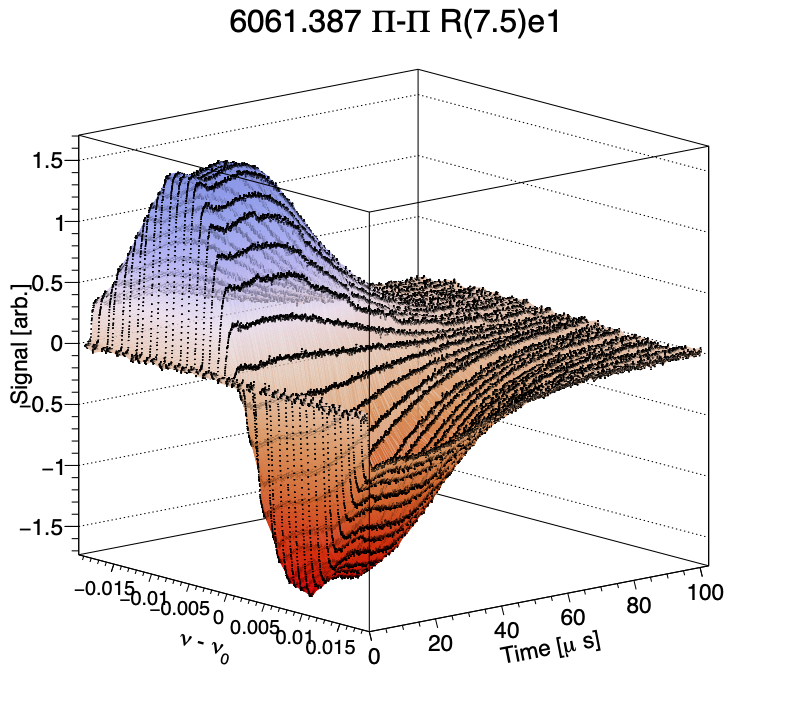}
    \caption{\footnotesize{}}
  \end{subfigure}%
  \caption{\label{fig:surfaces}\footnotesize{3-dimensional plots of signal strength against frequency and time for the two types of kinetic behaviour exhibited by the strong spectral lines.  Data were recorded at a total pressure of 500 mtorr in a 10\% mixture of \ce{CF3CCH} in argon.}}
\end{figure*}

The time-dependent behavior of the strong lines fell into two distinct categories which support the spectroscopic assignments.  The strong lines in the P- and R-branch pattern assigned to the $(6055.6)^2\Sigma \leftarrow (000)^2\Sigma$ transition share the same behavior, illustrated figure \ref{fig:surfaces}a, where the initial population, as evidenced by an initial instrument-limited rise, was small and the signal grew on a timescale of a few 10s of \us before decaying.  Based on earlier work in the group,\cite{LeISMS2017, Le2016} this kinetic behavior is characteristic of absorptions from the vibrationless state of the radical.  The growth comes from collisional relaxation from higher vibrations levels formed in the photolysis. \\

Meanwhile, the four-line pattern assigned to the $\Pi-\Pi$ transition, exhibited different behavior as shown in figure \ref{fig:surfaces}b.  An instrument-limited rise, interpreted as production of radical in the probed level as part of the nascent distribution is followed, at 500 mtorr pressure, by a period of approximately 10 to 20 \us where the signal exhibits a complex, pressure-dependent behavior, before a relatively slow decay.  The slow decay is similar to that seen for \vib{2}{0} levels and interpreted as chemical loss due to reaction of thermalized \ce{C2H} with the precursor.  Based on our earlier work\cite{Le2016, LeISMS2017}, this kinetics is characteristic of radicals in low bending vibrational levels, \textit{v}$_2$ = 1, or 2.  \\

\subsection{The CCH $6817.9\,^2\Pi$ state}
This state was identified as the upper state of a transition from the (000)$^2\Sigma$  ground state of CCH in recent work by Le et al.\cite{Le2018}  It is of interest for the present work because hot band lines from levels of $^2\Sigma$ and $^2\Delta$ symmetry in the bending overtone level (020) will occur in the region of the present spectrum.  These lower state levels are well known from laser induced fluorescence spectroscopy by Hsu et al.\cite{Hsu1993,Hsu1995,Chiang1999} and microwave studies by Killian et al.,\cite{Killian2007} so that the expected transitions in the present region should have been easily predicted.  \\

However, we discovered that the original analysis of the data suffered from some errors.  Firstly, the analysis code had a rotational constant factor missing from the diagonal elements expressed in the $\mathbf{N}^2$ representation, leading to a difference in the absolute state energy.  In addition, assignments of some transitions terminating in the F$_1$ levels were wrong because an error was made in the eigenvalue identification at low $J$, and the spectrum fitted as far as possible with that error.  Correcting these errors resulted in an approximately 1.4 \wn  shift in the upper-band origin to 6817.9 \wn, and altered the $\Lambda-$doubling parameters.  The Hamiltonian used in the present analysis is given in equations (\ref{equ:DPiMtx}) and (\ref{equ:DPiPty}). The complete dataset with the fit details is available in the supplementary data for this paper.  The newly determined parameters are given in the fourth column of table \ref{tab:upper_parms}.  The overall standard deviation of the fit relative to the estimated uncertainty of $\pm 0.002$ \wn for measured line positions, was 1.89, far better than previously, using fewer high-order $\Lambda-$doubling parameters.  The observation of a strong perturbation in the upper levels of $e-$symmetry with $J' > 23/2$ evident in the R-branch transitions, clear in figure 2 of \cite{Le2018} remains.  The P-branch lines accessing these levels are at frequencies below those scanned and cannot be used to confirm assignments of strongly perturbed levels.  \\

As noted previously,\cite{Le2018} Q-branch transitions accessing the other parity levels of the upper state cannot be assigned with complete confidence in a $^2\Pi\, -\, ^2\Sigma$ band because there are no allowed transitions to provide lower state combination differences to confirm them.  As before, the assignments were made on the basis of the smallest $\Lambda-$doubling and consistency with the rotational spacings in the molecule.  The $\Lambda-$doubling parameters are different to, and significantly better determined than, those reported previously.  \\

With the upper level energies now well determined, we found there were still systematic differences between the observed spectrum and transitions predicted for the $(6817.9)\,^2\Pi \,\leftarrow\,(020)\,^2\Sigma^+$ hotband transition.  The lower state energy level structure was also reported by Le et al.\cite{Le2018} from a combined fit of data previously published  at that time and combination differences from a new near-IR hotband, $^2\Sigma (7527.1) - ^2\Sigma (020)$. However, the upper state levels had some perturbations and these affected the global fit of the data including lower state rotational transitions.  In order to remove these effects, we have refit the microwave and appropriately weighted combination differences from all published data relating to the (020) $^2\Sigma^+$ level.  The overall standard deviation of the fit was 3.4 relative to the weights of the data, with the high-$J$ combination differences derived from the near-IR data sightly inconsistent with those derived from the UV data.\cite{Chiang1999}  All the observed microwave transitions were fit to the expected experimental uncertainties.\cite{Killian2007} The resulting parameters are given in the 4th column of table (\ref{tab:lower_parms}).  The dataset used and complete fitting results are available in the supplementary data for this paper.\cite{Supplemental}  \\

With these refined parameters for the states involved in the expected $(6817.9)\,^2\Pi \,\leftarrow\,(020)\,^2\Sigma^+$ hot band transition, we were able to identify a series of P- and Q-branch transitions that are expected to be the strongest in the band in the 6805 to 6830 \wn region.  Their positions were well predicted with a 0.049(12) \wn \, shift to higher energy.  This corresponds to shift of the same size, but opposite sign  in the $T_0$ value for the lower level, \textit{i.e.} $T_0((020)\,^2\Sigma^+)\,=\,794.251(12)$ from the present data, compared to that determined by Chiang et al.\cite{Chiang1999} quoted in table \ref{tab:lower_parms} of 794.300(6).  \\

For completeness, we have also refitted the previously published data for the $^2\Sigma^+(7527.1) -\, ^2\Sigma^+(020)$ band\cite{Le2018} using the refined lower level parameters in table \ref{tab:lower_parms}.  The results are given in table \ref{tab:7527parms}.  The parameters are slightly changed from previously. The quoted error in the $T_0$ value does not take into account the uncertainty in the lower state energy of at least $\pm0.006$ \wn, see above.  The overall RMS standard deviation of the fit is reasonable, given the weakness of many of the lines in this spectrum, see figure 1 of reference \cite{Le2018}.\\     

\begin{table}[t]
		\centering
		\caption{Spectroscopic parameters in wavenumbers (\wn) for the $^2\Sigma^+ (7527.1)$ level}
	\label{tab:7527parms}
  \begin{tabular}{ll}
		\hline
		\hline
		Parameter \,\, & $^{2}\Sigma ^{+}$ (7527.1)	\\
		\hline
		$B$                          & 1.4301257(15)$^a$\\
		$10^5\,D$                    & -8.0891(18) \\
		$10^9\,H$                    & -9.840(65) \\
		$10^2\, \gamma $             & -0.19279(19) \\
        Origin                       & 7527.10618(3) \\
        Fit RMS Dev.$^b$             & 4.6 \\
		\hline
		\hline
		\multicolumn{2}{l}{\textit{a.} This work. The numbers in }\\
		\multicolumn{2}{l}{parenthesis are one standard  }\\
		\multicolumn{2}{l}{deviation of the fit in units} \\
		\multicolumn{2}{l}{ of the last quoted decimal place.} \\
        \multicolumn{2}{l}{\textit{b.} Standard deviation of the } \\
		\multicolumn{2}{l}{fit relative to estimated} \\
        \multicolumn{2}{l}{measurement uncertainty of 0.004 \wn.}
        \\	
  \end{tabular}
\end{table}
Transitions from the $\tilde{X}(020)\,^2\Delta$ level to the same $^2\Pi (6817.9)$ upper level should also appear in the region studied in the present work, assuming the transition moments of the two vibrational levels are similar, but none was identified.  The greater number of rotational levels in the $^2\Delta$ component leads to a considerable dilution in the average line strength compared to the corresponding $^2\Sigma$ hotband if the vibronic transition moments are the same, so it is perhaps not surprising the transitions were not seen in this work.

\section{Conclusions}
We have reported new measurements of the near infrared spectrum of the chemically important ethynyl radical. The new data have been analyzed together with data from previously reported measurements by ourselves and others to build up a coherent picture of the rotational, fine, and hyperfine energy structure of the radical. The strong hot band spectrum lies in a convenient region of the spectrum for modern diode and fiber laser sources and detectors, potentially making it an ideal vehicle for future analytical applications.  


\section*{Acknowledgements}

This work was supported by the U.S. Department of Energy, Office of Science, Division of Chemical Sciences, Geosciences and Biosciences within the Office of Basic Energy Sciences, under Award Number DE-SC0018950.  The experimental measurements were carried out at Brookhaven National Laboratory under Contract No. DE-SC0012704 with the U.S. Department of Energy, Office of Science, and supported by its Division of Chemical Sciences, Geosciences and Biosciences within the Office of Basic Energy Sciences. 

\section*{Supplementary Data}
Supplementary Data associated with this work consists of three files:\\
1. A spreadsheet containing multiple worksheets that detail the assigned data and the least squares fits to obtain the molecular parameters in tables of this manuscript.\\
2. A text file containing the results of the fit to the (010) $^2\Pi$ energy levels. \\
3. A zipped comma-separated-variable file containing the spectrum plotted in Figure (1) of this paper.

\small
\bibliography{CCH6040.bib}


\end{document}